\documentclass[twoside]{article}
\usepackage[latin1]{inputenc}
\usepackage{t1enc}
\usepackage{a4}
\usepackage{tabularx}
\usepackage{epsf}
\usepackage{astron}   

\newcommand{\hb}{H$\beta$}
\newcommand{\mgb}{Mg$_b$}

\begin{document}
\bibliographystyle{astron}   

\setcounter{figure}{0}
\setcounter{section}{0}
\setcounter{equation}{0}
\renewcommand{\refname}{}

\begin{center}
{\Large\bf
Evolution of Early--type Galaxies in Clusters\\[0.2cm]}

Bodo L. Ziegler \\[0.17cm]
Universit\"atssternwarte G\"ottingen \\
Geismarlandstr. 11, D--37083 G\"ottingen, Germany \\
bziegler@uni-sw.gwdg.de, http://www.uni-sw.gwdg.de/~bziegler
\end{center}

\vspace{0.5cm}

\begin{abstract}
\noindent{\it
  This is a short review on the evolution of elliptical and lenticular/S0
  galaxies in the dense environment of rich clusters for look-back times of
  more than half the age of the Universe. In addition, new data are presented
  for the cluster Abell\,2218. 
  
  Scaling relations like the Fundamental Plane provide strong evidence that
  luminous ellipticals evolve purely passively up to redshifts of $z=0.8$
  implying that no significant new stellar population was created during that
  time. This picture is supported by tight Mg--$\sigma$ and colour--magnitude
  relations at various redshifts and age/metallicity diagnostic diagrams based
  on absorption lines. On the other hand, there is a strong decline in the
  frequency of S0 galaxies with redshift as was revealed by HST images of
  distant rich clusters. In the same clusters, however, a significant
  proportion
  of the galaxies have disturbed morphologies and signs of ongoing
  interaction/merging.  Numerous post--starburst galaxies (like E$+$A galaxies)
  are found as well as passive spiral galaxies, while these galaxy types are
  less present in local clusters. This raises the question whether the
  spiral galaxies falling into the cluster from the surrounding field
  could be transformed into S0 galaxies. A possible scenario would be that the
  intracluster medium first deprives the spirals
  of all their gas leading to the end of star formation. The subsequent
  evolution within the tidal field of the cluster could then turn the spirals
  into S0 galaxies over the ages.

  The study of a great number of early--type galaxies per cluster including 
  sub--L$^*$ galaxies reveals, however, that this picture needs some finetuning
  since the relations mentioned above have a larger spread than those based on
  the observation of just a few luminous galaxies. First results are presented
  for 48 early--type galaxies in the cluster Abell\,2218 at a redshift of
  $z=0.18$. }
\end{abstract}

\section{Introduction}
\label{intro}
In the morphological scheme of Hubble's ``Tuning Fork'', the branch of 
early--type galaxies comprises the elliptical galaxies and the lenticular or S0
galaxies, which have a bulge and a disk component \cite{Hubbl36}. Ellipticals
can be further divided into boxy and disky ellipticals according to their
isophote shapes \cite{KB96}. Since disky ellipticals contain small stellar 
disks they may form a continuous sequence with the S0 galaxies. In a 
dynamical view,
rotational support decreases from S0 to disky ellipticals whereas the random
motion of the stars, measured by the velocity dispersion $\sigma$, gets more 
and more
important. Another common feature of early--type galaxies is the lack of any
significant amount of cold gas on global scales so that there is no significant
ongoing star formation.

The currently favoured theory that combines cosmology and galaxy formation is
the Cold Dark Matter model, which comes in various flavours. It predicts that
galaxies merge hierarchically, i.e., smaller entities hit
one against the other and gradually build up a large galaxy; the so-called
mergertree scenario \cite{LC93}. Especially in the
field, merging disks get heated up until they form a spheroidal, but at a later
stage a disk can be formed again when the evaporated gas, which was trapped in
the dark halo, is falling back onto the galaxy \cite{BCF97}. In this way, the
mass of a galaxy is increased in jumps each time a merger occurs and a new
stellar population may be created which dilutes the underlying older population
and makes the mean stellar age younger. This is described by semi--analytic 
\cite{KC98} and hydrodynamic SPH \cite{SN99}
models, which are able to incorporate star formation and feedback processes.

 
On the other hand, stellar population models, which predict the spectra and
hence the colours and absorption lines of a galaxy for all time steps of its
evolution based on the isochrones of the stellar content, assume just a
certain time dependent star formation rate for a single metallicity and initial
mass function. Most of them have not yet incorporated the superposition of 
different stellar populations having various ages and metallicities as the 
hierarchical merging picture implies (but see e.g. Sansom and Proctor 
\cite*{SP98b}). Nevertheless, these passive evolution models are able to 
reproduce the observed
properties of both local and distant galaxies (e.g., Bruzual \& Charlot,
1993)\nocite{BC93}. The simplest model picks up a short burst of star formation
in which all the stars of a galaxy are created, the so-called Simple Stellar 
Population model (SSP). It describes the observed evolution of early--type 
galaxies quite well.
The main features are the rapid decline of the blue luminosity so that the red
light is dominating after about 3\,Gyrs, the rapid diminishing of the Balmer
lines and the strengthening of the metal lines on the same timescale.  However,
stellar populations are degenerate between age and metallicity. Thus, the 
spectra of SSP's are nearly identical if the age is increased
by a factor of 3 and, simultaneously, the metallicity decreased by a factor of 2
\cite{Worth94}. For example, a spectrum at an age of 15\,Gyrs and a
metallicity of [Fe/H] $=-0.1$ looks the same as that at an age of 5\,Gyrs and a
metallicity of [Fe/H] $=+0.2$. This means that the colours and most absorption
lines are degenerate, too, and can not be used to distinguish between age and
metallicity effects. Only a few indexes like the Balmer lines are more 
dependent on age and the combinations of some metal lines like [MgFe] are more
metallicity sensitive, so that they can be evaluated in an age/metallicity 
diagnostic diagram.

\begin{figure}[htb]
\centerline{\epsfxsize=10cm \epsfbox{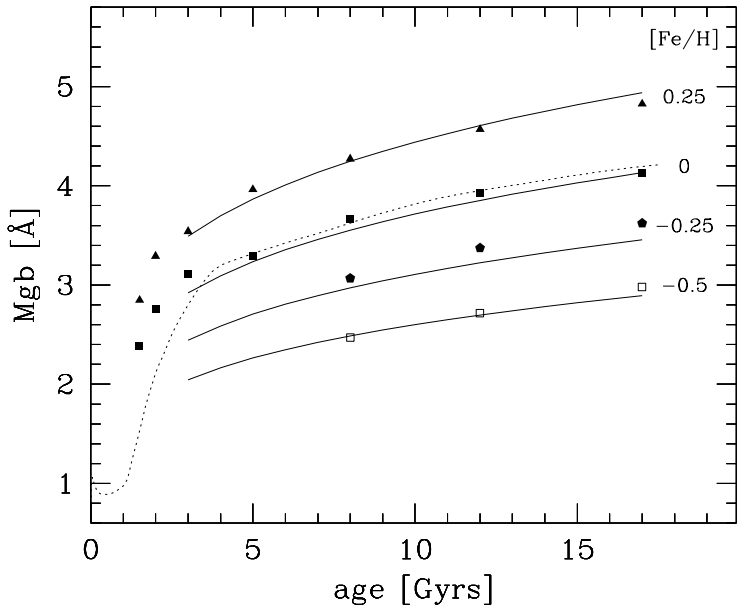}}
\caption[]{The dependence of the \mgb\ absorption index on age and metallicity
(from Ziegler \& Bender, 1997). Symbols are model data from Worthey (1994), 
the dotted 
line represents the SSP model of Bruzual \& Charlot for solar metallicity and
the solid lines follow Equation\,\ref{mgbtZeq}.}
\label{mgbtZ}
\end{figure}
\nocite{ZB97}
 
\section{Evolution of elliptical galaxies}
\label{e_evol}
\begin{figure}[hbt]
\centerline{\epsfxsize=10cm \epsfbox{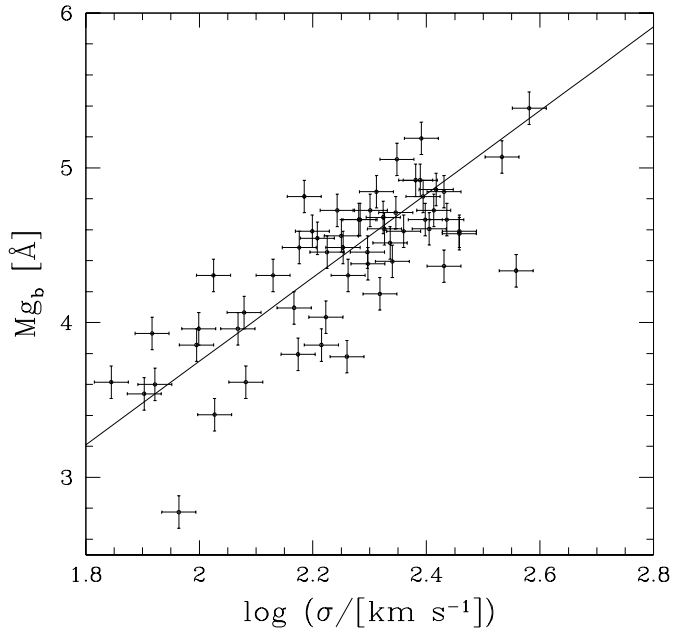}}
\caption[]{The \mgb--$\sigma$ relation of elliptical galaxies in the local 
Virgo and Coma clusters (data from Dressler et al., 1987, as transformed like
in Ziegler \& Bender, 1997). }
\label{mgbs}
\end{figure}

First evidence for an old age of ellipticals in clusters came from the
observations of very tight colour--magnitude \cite{BLE92}, Fundamental Plane
\cite{BBF92} and Mg--$\sigma$ \cite{BBF93} relations in local clusters. The 
\mgb\ absorption index ($\lambda_0\approx5172$\AA) follows Worthey's 3/2--rule
(see Fig.\,\ref{mgbtZ}) and
from his SSP models the following time and metallicity dependence can be
derived: 
\begin{equation}
\label{mgbtZeq}
\Delta \log \mbox{Mg}_b = 0.20 \Delta \log t + 0.31 \Delta \log Z/Z_o
\end{equation}
The observed narrow spread in the \mgb--$\sigma$ relation of ellipticals in the
Virgo and Coma clusters (see Fig.\,\ref{mgbs}), therefore, translates into 
small limits for the dispersion in age and metallicity at fixed velocity 
dispersion\cite{CBDMSW99}. For the most massive ellipticals in Coma, for
example, relative age and relative metallicity are constrained to 17\% and 
11\%, respectively \cite{ZB97}.
This means in the picture of passive evolution (i.e. no merging), that when 
\mgb\ line strengths of distant and, therefore, younger
galaxies are compared with those of nearby ones at the same velocity 
dispersion $\sigma$, any difference can mostly be attributed to an age effect.
In this sense, the \mgb--$\sigma$ relation breaks the age/metallicity 
degeneracy and is a powerful tool to investigate the evolution of elliptical
galaxies. 

\nocite{DLBDFTW87}

\begin{figure}[hbt]
\centerline{\epsfxsize=10cm \epsfbox{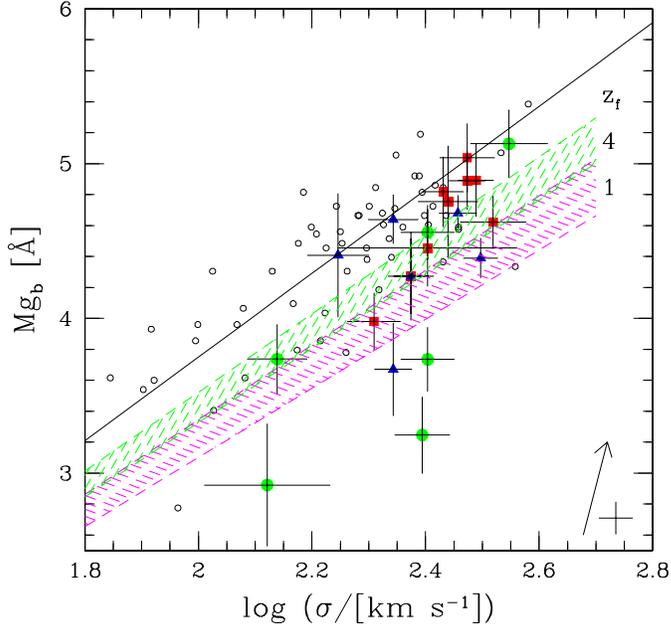}}
\caption[]{The \mgb--$\sigma$ relation of elliptical galaxies in three distant
clusters at $z\approx0.4$ (points with errorbars) compared to local cluster
ellipticals (from Ziegler \& Bender, 1997). Hatched areas represent the 
expected range of \mgb\ line strengths for redshifts of formation $z_f$ of 1 
and 4 for a variety of cosmological parameters. At the lower right corner, the
arrow represents the aperture correction applied to the distant data points
and the cross indicates the average error of the measurements of the local
galaxies.}
\label{mgbsz}
\end{figure}

This age determination technique was first applied by us to early--type 
galaxies in three clusters at a redshift of $z\approx0.4$: Abell\,370 
($z=0.375$), MS\,1512$+$36 ($z=0.372$) and Cl\,0949$+$44 ($z=0.377$)
\cite{BZB96,ZB97}. 28 nights
at the Calar Alto 3.5m-telescope, 9 nights at the ESO NTT and 2 nights at the
ESO 3.6m-telescope were spent to get 21 good quality spectra. Except from a few
outliers (probably E$+$A galaxies), the reduction in the \mgb\ absorption is 
low and the mean offset from the local fit line is about 0.4\AA\ (see
Fig.\,\ref{mgbsz}). This means, that the distant galaxies must be quite old
themselves, because the lookback time of about 5\,Gyrs did not produce a higher
difference in \mgb\ and that they indeed evolved only passively. Assuming the 
metallicity did not change at all during that time,
Equation\,\ref{mgbtZeq} can be transformed to predict the \mgb\ absorption for
any given age. It depends, though, on the redshift of formation $z_f$ and the
cosmological parameters. In Fig.\,\ref{mgbsz}, hatched areas indicate the 
expected location of the 
\mgb\ absorption for two different
formation redshifts and a range of plausible values of $H_0$ and $q_0$.
The comparison with the data points leads to the conclusion, that the
observed galaxies must have formed at $z_f>2$, the most massive ones even at
$z_f>4$. But only the bulk of the stars must have come into existence at these
epochs, the galaxies themselves could have been assembled at a later time if no
new star formation took place during that process. These observations are still
in agreement with CDM models, since these predict, that the last major merger 
of a galaxy in the cluster environment occurs at about $z=2$ \cite{Kauff96}.

\begin{figure}[hbt]
\centerline{\epsfxsize=9.5cm \epsfbox{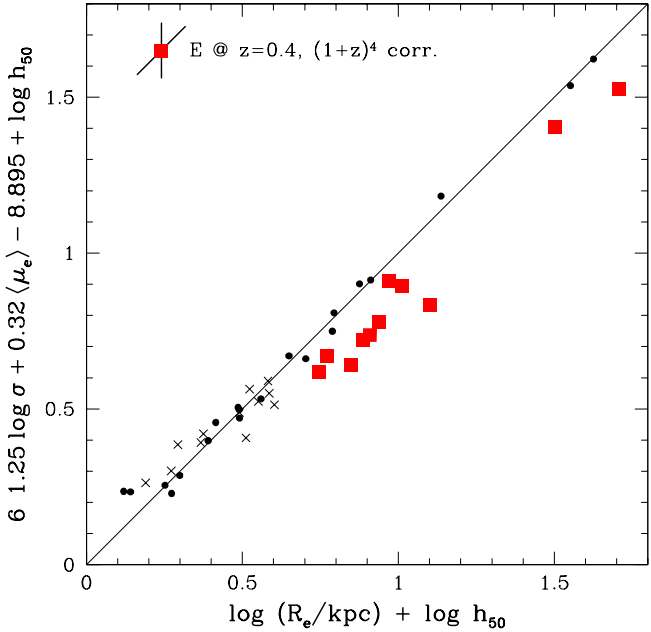}}
\caption[]{The Fundamental Plane seen edge-on for Coma ellipticals (small 
filled circles) and S0 galaxies (crosses) compared to early--type galaxies in
three clusters at $z\approx0.4$ (big filled squares) (adapted from Bender et
al., 1998). The observed HST magnitudes of the distant galaxies were 
transformed to restframe $B$ and corrected for surface brightness dimming. A
typical error is plotted in the upper left corner.}
\label{fp}
\end{figure}

Another physical relationship of early--type galaxies is the Fundamental 
Plane (FP), which is set by the three observables velocity dispersion $\sigma$,
effective or half--light radius $R_e$ and the mean surface brightness 
$\langle\mu_e\rangle$ within $R_e$. It is a consequence of the virial theorem
and a slight dependence of the mass-to-light ratio on the mass (e.g.
Bender et al., 1992). Seen edge-on, the FP is very tight and both elliptical 
and S0 galaxies in the Coma cluster follow the same fit line although they are
dynamically different \cite{BSZBBGH97b}. From the HST images of the three
distant clusters we were able to derive accurately the photometric parameters
\cite{SMGBZ00}. 
Transforming the observed magnitudes into restframe $B$ magnitudes, they can be
compared to the Coma galaxies (see Fig.\,\ref{fp}) and a mean brightening of
about 0.5 mag for the distant galaxies is derived. 

In the meanwhile, other groups have used the FP as well to study the evolution
of early--type galaxies in clusters at different redshifts: Abell\,2218 at
$z=0.175$ \cite{JFHD99,ZBSDL99}, Abell\,665 $z=0.181$ \cite{JFHD99}, 
Abell\,2390 $z=0.24$ \cite{ZBSD00}, Cl\,1358$+$62
$z=0.33$ \cite{KDFIF97}, Cl\,0024$+$16 $z=0.39$ \cite{DF96}, MS\,2053--04
$z=0.58$ \cite{KDFIF97} and MS\,1054--03 $z=0.83$ \cite{DFKI98}. The combined
results strongly favour a purely passive evolution of the stellar populations
both in surface brightness and mass-to-light ratio (see for a summary
J{\o}rgensen et al. \cite*{JFHD99}). For open and $\lambda$--dominated 
cosmologies
and a Salpeter IMF \cite{Salpe55}, redshifts of formation greater than 2 are
required to yield the corresponding old ages \cite{BSZBBGH97b,DFKI98}.

\clearpage

\section{Evolution of S0 galaxies}
\begin{figure}[hbt]
\centerline{\epsfxsize=9.5cm \epsfbox{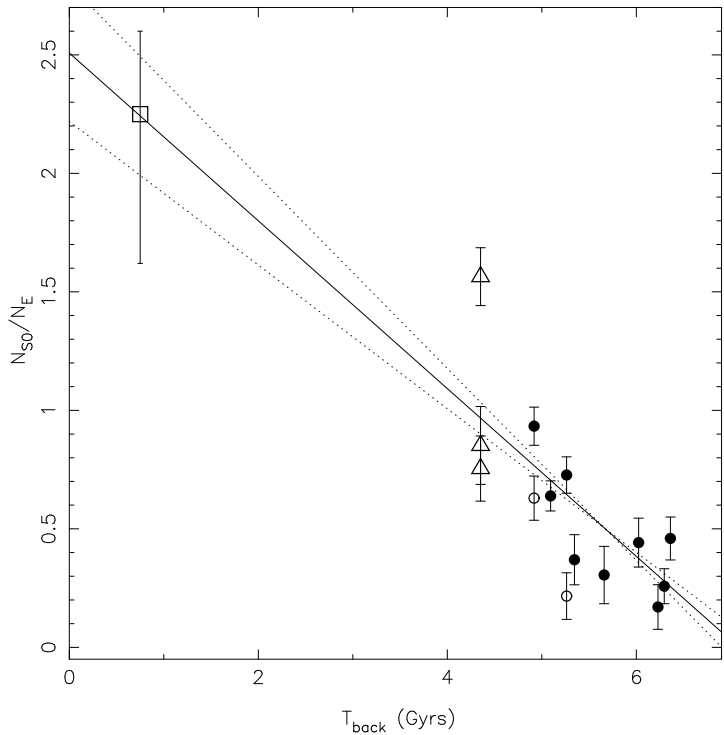}}
\caption[]{The proportion of bright S0 to elliptical galaxies in rich clusters
 as a function of lookback time (adapted from Dressler et al., 1997). The fit
 shown (with $\pm1\sigma$ limits as dotted lines) was made to the MORPHS data
 (filled circles, Smail et al., 1997) and is compatible with clusters at 
$z=0.3$ (open triangles, Couch et al., 1998) and the local reference (Dressler
1980). The 2 open circles represent the outer regions of 2 clusters, which are
not directly comparable to the other points.} 
\label{s0evol}
\end{figure}

In contrast to the elliptical galaxies, S0 or lenticular galaxies show a
dramatic evolution with lookback time in the dense environment of rich 
clusters.
But this was only recently discovered by images from the \textit{Hubble Space
Telescope}, whose \textit{WFPC2} camera made it possible to resolve even the
faint galaxies in distant clusters and determine their morphology (e.g.,
Dressler et al. \cite*{DOCSE97}). Apart from the dwarf galaxies, the S0 
galaxies form the dominant population in local rich clusters, followed by the
ellipticals, whereas spirals contribute roughly only 10\%. This situation is
very different in rich clusters at a redshift of $z\approx0.5$. There, spiral
and disturbed galaxies make up the major part of the luminous galaxies, whereas
S0 galaxies are hardly found (10--20\%), see Fig.\,\ref{s0evol}.

\nocite{Dress80,CBSES98,SDCEOBS97}

\vspace*{2cm}

On the other hand, Butcher \& Oemler found a rapid increase in the ``blue''
galaxy population in rich clusters with redshift already twenty years ago
\cite{BO78a,BO84a}. On average, the fraction of ``blue'' galaxies rises from
0.05 in local clusters to 0.25 in clusters at $z=0.5$. Early spectroscopy of
such BO--galaxies revealed in some cases a spectral energy distribution similar
to normal ellipticals but with a stronger absorption of the Balmer lines, so
that they were dubbed E$+$A galaxies (e.g. Butcher \& Oemler \cite*{BO84b}). 
The idea behind that was that these post--starburst galaxies were ellipticals 
who had undergone a second burst of star formation in the recent past
($\leq$2\,Gyrs), so that the newborn $A$ stars were still responsible for the
increased strength of the Balmer lines. But using high--resolution HST images,
stellar disks were detected in most of these E$+$A galaxies indicating that
spirals are the more probable progenitors rather than ellipticals (e.g., Couch
et al., 1998). New spectroscopic observations have found in addition a
number of dust--enshrouded starburst galaxies and a population
of red, passive spiral galaxies with no significant emission 
lines \cite{PSDCB99}.

\begin{figure}[hbt]
\centerline{\epsfxsize=8.5cm \epsfbox{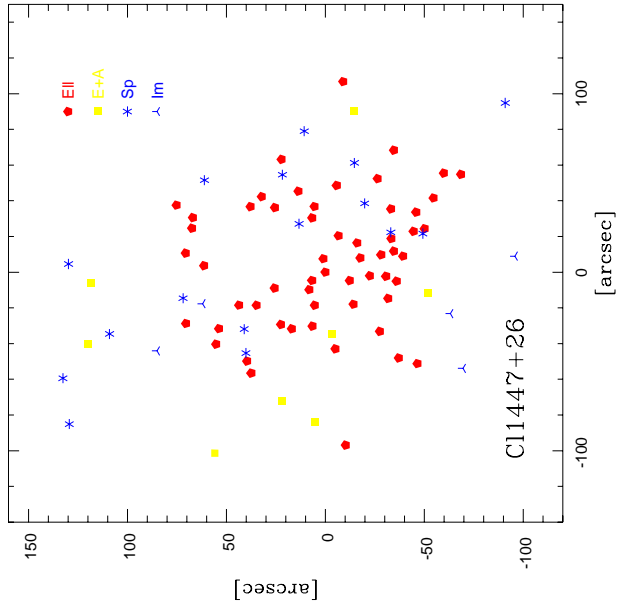}}
\caption[]{Radial distribution in the cluster Cl\,1447$+$26 at $z=0.37$ (from
Belloni et al., 1997).
Early--type galaxies (diamonds) are concentrated to the core, whereas spirals
(stars), irregulars (crosses) and E$+$A galaxies (box) are located in the outer
regions. } 
\label{cl1447}
\end{figure}

Many other galaxies in the distant clusters have a
disturbed morphology and some show clear signatures of past or ongoing
interaction or merging \cite{DOSL94,SDCEOBS97}. The majority of the E$+$A 
post--starburst galaxies are located at the rim of a cluster if it is regular 
or in between subcomponents \cite{BVR97,BMYCE97}. In these regions, 
most of the spiral and irregular galaxies 
are found, too, whereas the early--type galaxies are concentrated to the core 
like in local clusters, see Fig.\,\ref{cl1447}. S0 galaxies themselves were
found to be bluer at the outskirts of a cluster than in the core \cite{DFK98}.

All these findings give clear evidence that galaxy transformation is a common
phenomenon in clusters so that the morphology of a galaxy can be changed. There
are signs that galaxies interact with each other and even merge together.
But the high velocity dispersion in rich clusters makes the effective cross 
section for such encounters small, so that the influence of the hot 
intracluster medium is much more important. This ICM was already detected by 
the first X--ray
satellites, since its temperature is between 1 and 10 million degrees. In 
nearby clusters, H\textsc{i} studies revealed spiral galaxies deficient in 
H\textsc{i} and a displacement of the H\textsc{i} disk from the stellar disk, 
which is the larger the closer the galaxy is located to the center 
\cite{VJ91,CKBG94,BCBG98}. This could be caused by ram--pressure stripping of 
the ICM or tidal interactions between the galaxies.

\begin{figure}[hbt]
\centerline{\epsfxsize=10cm \epsfbox{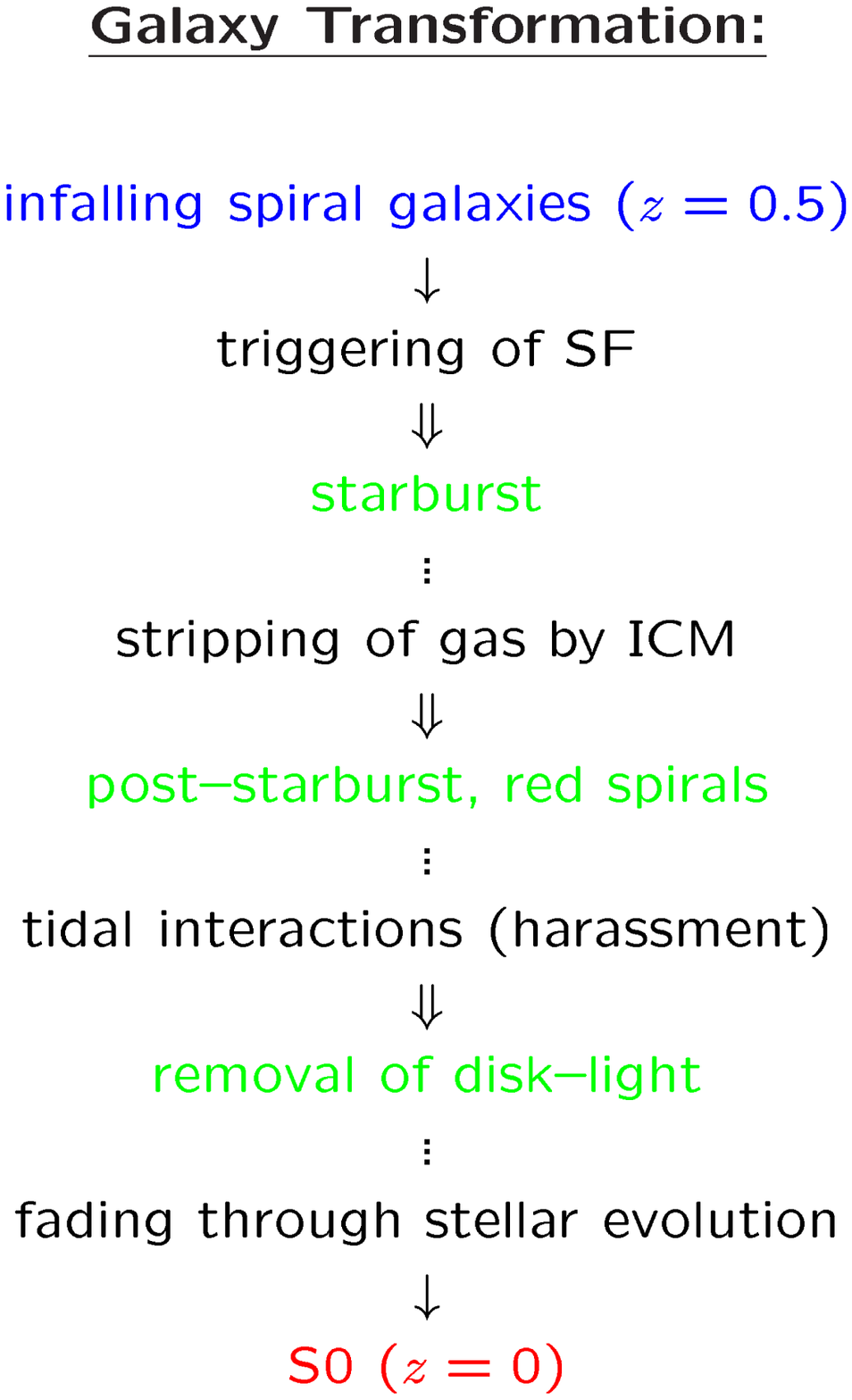}}
\caption[]{Flow chart of a scenario for the transformation from spirals into S0
galaxies in rich clusters.} 
\label{gt}
\end{figure}

Which mechanisms lead to the transformation into S0 galaxies is still unknown,
so that the strong increase in the frequency of S0 galaxies during the past
5\,Gyrs can not yet be explained. Spectroscopic observations have ruled out a 
simple fading of spiral galaxies \cite{CS87,BMYCE97,DFK98}. The strengths of 
the Balmer absorption lines in some galaxies require that a truncation 
mechanism
must act to cut off star formation very quickly and must also promote
bursts of star formation \cite{BAECSS96}. An appealing scenario is that spiral
galaxies falling into the cluster from the surrounding field undergo a dusty
starburst phase. This 
increased activity causes the Butcher--Oemler effect. Ram--pressure stripping 
by the ICM then leads to a wide spread and rapid decline in the star formation 
rates resulting into the post--starburst galaxies and red, passive spirals. 
However, other processes may be equally important, such as tidal stripping 
 \cite{GTC96} or galaxy--galaxy interactions \cite{LH94}.
The timescale for the transformations described thus far is rather short 
($<$1\,Gyr, e.g., Barnes and Hernquist \cite*{BH92}). The subsequent
evolution takes much longer and the permanent influence of the tidal field of
the cluster as a whole (harassment) may cause a substantial removal of disk 
light \cite{MLKDO96}. The end product of this scenario may then be an S0
galaxy (see the flowchart, Fig.\,\ref{gt} and Poggianti et al. 
\cite*{PSDCB99}).

\clearpage

\section{Evolutionary status of early--type galaxies in Abell\,2218}

The determination of morphology is increasingly difficult the more distant
and, therefore, fainter the galaxies are, even on HST images \cite{SDCEOBS97}.
For example, the average size of L$^*$ galaxies at $z=0.5$ is only 
$\approx1-1.5$\,arcsec
\cite{ZSBBGS99}. But even in closer systems it is hard to detect disks and 
their visibility
depends strongly on the inclination. A round S0 galaxy seen face-on may be 
classified as an elliptical. The MORPHS group has estimated that about 15\%
of the S0 population may be misclassified as ellipticals independent of
redshift. An additional problem of the past studies is the
low number of observed galaxies. In most cases, spectra of only a handful or a
dozen of the brightest cluster members were taken. Therefore, the evolution 
only of the more massive galaxies have been studied in FP analyses.

\begin{figure}[hbt]
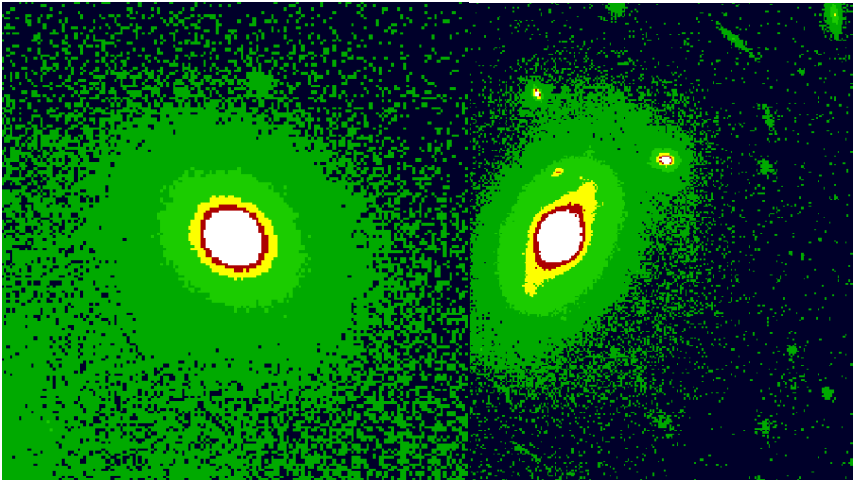

\centerline{\epsfxsize=6.2cm \epsfbox{e.cps} \epsfxsize=5.1cm \epsfbox{s0.cps}}
\caption[]{Bona-fide examples of an elliptical and an S0 galaxy in the cluster
Abell\,2218 ($z=0.175$) extracted from an HST/WFPC2 image in the F702W 
filter. More examples can be found in Ziegler et al. 2000a.} 
\label{ES0}
\end{figure}

To overcome this bias problem, we started a project in 1997 to observe for the
first time a great number of early--type galaxies in the cluster Abell\,2218
($z=0.175$). Multi--object spectroscopy was done using the LDSS2 spectrograph
at the William--Herschel telescope with three different masks \cite{ZBSDL99}.
Galaxies were selected by a colour criterion and encompass several sub-L$^*$
galaxies. The total exposure time of $\approx5$ hours allowed the extraction 
of
high signal-to-noise spectra of 48 different early--type galaxies. The average
size of these galaxies is $\approx2-4$\,arcsec, which allows the accurate
determination of the photometric parameters of the subsample of 20 galaxies
contained in HST images. For the other galaxies they were measured from $UBVI$
photometry obtained at the Palomar 5m-telescope. In the HST subsample, there 
are 9 S0/SB0 galaxies making up 50\% and 2 early spirals.  Fig.\,\ref{ES0} 
shows bona-fide examples of  an elliptical and an S0 galaxy. Five of the 
galaxies outside
the HST field have clear evidence for a disk component, too.

\begin{figure}[hb]
\centerline{\epsfxsize=10cm \epsfbox{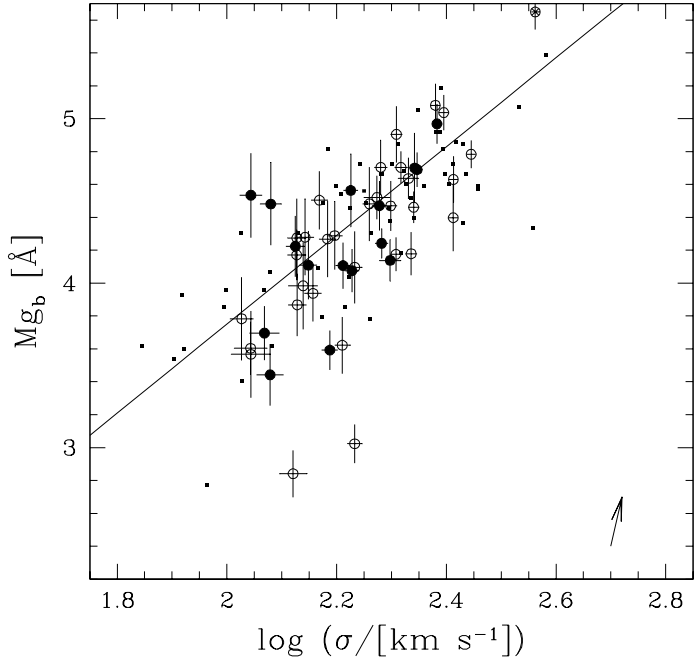}}
\caption[]{The \mgb--$\sigma$ relation of early--type galaxies in the
cluster Abell\,2218 at $z=0.175$ (points with errorbars) compared to 
ellipticals in the local Coma and Virgo clusters  (from Ziegler et al., 
2000a). S0 galaxies are denoted by filled circles, ellipticals by open 
circles. The crossed circle marks a galaxy probably undergoing a merger.
The arrow in the lower right corner represents the aperture correction applied
to the A\,2218 data points. The line represents the local \mgb--$\sigma$ 
relation.}
\label{mgbs_a}
\end{figure}

The velocity dispersion $\sigma$ and the line strengths of several absorption
lines were measured like for the cluster galaxies at $z\approx0.4$ (see
Section\,\ref{e_evol}). Fig.\,\ref{mgbs_a} compares the \mgb--$\sigma$ relation
of A\,2218 to the same local sample as in Figs.\,\ref{mgbs} and \ref{mgbsz}.
The spread of the distant relationship is almost the same as of the local one
and the zeropoints are nearly identical within the errors indicating a very old
age for the early--type galaxies in A\,2218. The slopes are also not very
different from each other, so that the distant and local galaxy samples could
have been drawn from the same population (KS statistics: $p=0.016$). 
But the most striking result is that
elliptical and S0 galaxies are equally distributed along the \mgb--$\sigma$ 
relation (KS statistics: $p=0.61$) so that there exist S0 galaxies which
are as old as ellipticals at a lookback time of $\approx2.5$\,Gyrs for A\,2218.
The two galaxies with very low \mgb\ absorption have rather high \hb\ values,
and could, therefore, be post--starburst galaxies like E$+$As. It is not
possible to detect any disk in these two galaxies in the ground--based images
(FWHM $\geq1.1$\,arcsec). Note also, that the two early spirals do not deviate
significantly from the \mgb--$\sigma$ relation.

\begin{figure}[htb]
\centerline{\epsfxsize=10cm \epsfbox{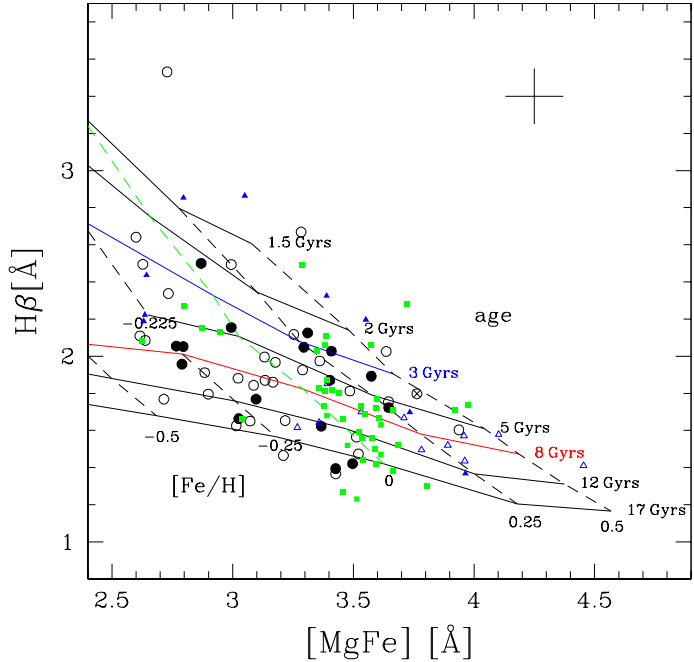}}
\caption[]{A line diagnostic diagram to disentangle age and metallicity effects
(from Ziegler et al., 2000a). See text for explanation. S0 galaxies in 
A\,2218: filled circles, ellipticals in A\,2218: open circles, a typical
error is shown in the upper right corner.
The underlying grid of lines with constant age (solid; 1.5 -- 17Gyrs)
and constant metallicity (dashed; [Fe/H]$=-0.5-+0.5$) was derived from 
SSP models by Worthey (1994). For
comparison, ellipticals (open triangles) and S0 galaxies (filled triangles)
of the local Fornax cluster (Kuntschner \& Davies, 1998) are plotted as well 
as Gonzalez' sample (1993) of nearby field and Virgo ellipticals (boxes).}
\label{mgfehb}
\end{figure}

Although most of the absorption line indexes of SSP models are degenerate in 
age and metallicity (see Section\,\ref{intro}), it is possible to construct 
line diagnostic diagrams. The reason is that some lines are more sensitive to
age others more to metallicity. The Balmer lines like \hb, e.g., depend very
much on age, whereas the combination between \mgb\ and Fe\,5270, [MgFe], is a 
measurement of the metallicity \cite{Worth94}. Such a diagram is shown in
Fig.\,\ref{mgfehb}. The galaxies in A\,2218 are spread out in age and
metallicity with ellipticals and S0s being almost
equally distributed. In particular, there are S0 galaxies whose mean
age can be fitted both by models for a very old age and by a very young age.
This is in contrast to the trend in the local poor Fornax cluster, where the
lenticulars are more concentrated towards younger ages \cite{KD98}. There
also seems to be a lack of very old and high metallicity galaxies in A\,2218.
But these galaxies are also missing in a sample of randomly selected nearby 
field and Virgo cluster early--type galaxies \cite{Gonza93}.

\begin{figure}[hbt]
\centerline{\epsfxsize=9.5cm \epsfbox{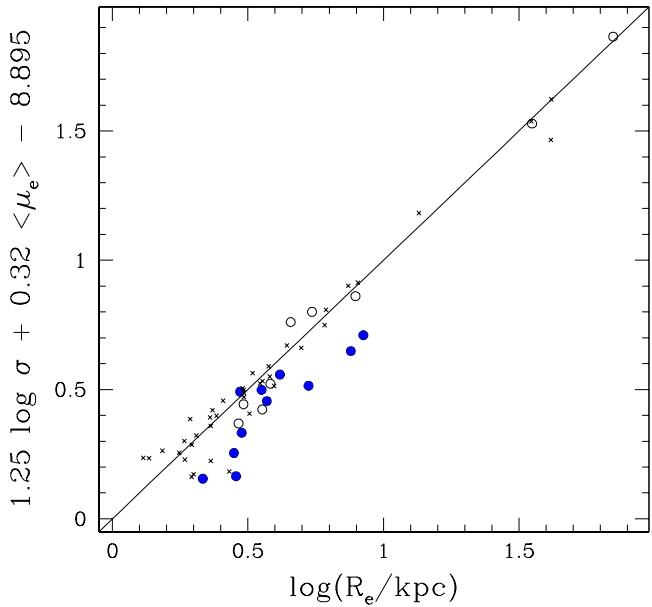}}
\caption[]{The Fundamental Plane seen edge-on for Coma early--type galaxies
(crosses, fit line) compared to ellipticals (open circles) and lenticulars 
(filled circles) in Abell\,2218 at $z=0.175$ (from Ziegler et al., 2000a).
The observed HST magnitudes of the distant galaxies were 
transformed to restframe $B$ and corrected for surface brightness dimming.} 
\label{fp_a}
\end{figure}

We have determined the structural ($R_e$) and photometric parameter 
($\langle\mu_e\rangle$) of the Fundamental Plane for the 20 early--type 
galaxies visible on the HST image of A\,2218 according to Saglia et al. 
\cite*{SBBBCDMW97}. In Fig.\,\ref{fp_a}, we compare the edge-on projection of
the FP with that of Coma early--type galaxies (cf. with Fig.\,\ref{fp}). The 
scatter is marginally larger than for Coma, and the slopes are not 
significantly different. There
seems to be some difference between elliptical and S0 galaxies with the latter
being brighter on average. If all of the offset is attributed purely to 
changes in luminosity, then the brightening is marginally consistent with pure
passive stellar population models (e.g. BC98 models, cf. Bruzual and Charlot
\cite*{BC93}).

\section{Summary and future}
In this short review it was shown, that luminous elliptial galaxies in rich
clusters do evolve just passively according to the fading of their stellar
content. The bulk of the stars must have been formed at high redshifts 
($z_f>2$), which is in line of predictions from CDM models in these dense
environments. Most of the evidence comes from Fundamental Plane analyses of
several clusters at redshifts from $z=0-0.8$, which is augmented by the study 
of \mgb--$\sigma$ relations and age/metallicity diagrams. 

In contrast to this smooth evolution, lenticular/S0 galaxies show a dramatic
evolution in their frequency. While they form the dominant population of the
bright galaxies in local rich clusters, their number diminishes to a few 
percent within a lookback time of about 5\,Gyrs. To explain this phenomenon, a
scenario was issued, in which spiral galaxies coming from the field 
surrounding a cluster are gradually transformed into S0 galaxies. An arriving
galaxy does not change its morphology immediately, but first its star 
formation is truncated due to the quick removal of the gas,
most probably by the ICM.
Subsequently, the permanent tidal interaction with the cluster potential and
other galaxies may change its appearance (bulge-to-disk ratio). Together with
the fading of the stellar population these processes may then transform the
spiral into an S0 galaxy as seen today.

The observation of a large sample of early--type galaxies in the cluster
Abell\,2218 at $z=0.18$ revealed that a refined picture of galaxy evolution is
necessary. Half of the 50 galaxies, which are made up by both luminous and 
sub--L$^*$ galaxies, are classified as ellipticals, the other half as
lenticulars. Both galaxy types are almost equally distributed along the
Fundamental Plane and \mgb--$\sigma$ relations. Modeling of the age sensitive
Balmer lines showed that the mean age of the stellar population of both
ellipticals and S0 galaxies varies between 2 and 20\,Gyrs in A\,2218.

\bigskip

Galaxy evolution certainly is science at the very frontier of astronomy and the
determination of the kinematics of faint galaxies is only possible with
observations at 10m--class telescopes. With the advent of the VLTs we embarked
on several projects aimed to further disentangle the phenomena connected to
galaxy evolution. These projects encompass the Fundamental Plane and 
\mgb--$\sigma$ analysis of
about 15 early--type galaxies in the cluster MS2053--04 at $z=0.6$ as well as a
detailed study of late--type galaxies in 6 different rich clusters at
$z=0.3-0.6$. Not only the star formation history of the galaxies located 
between the core and the very edge of the clusters, but also rotation curves
and, thus, the mass and dynamics will be measured. As a complement, the same
science will be made with a large sample of 170 field galaxies selected from
the FORS Deep Field. The combined results will put strong constraints on CDM 
models. 

To get a grip on 
the problem of galaxy transformation within clusters, we also started an 
intense campaign to observe poor X--ray detected clusters at $z=0.2-0.3$ with
ground--based optical (Calar Alto, HET, GEMINI) and radio telescopes (VLA) as
well as with HST and XMM.
The effectiveness of the competing processes invoked to explain galaxy
transformation varies according to density and virial temperature of the
local environment. For example, galaxy mergers are most effective in
lower galaxy density environments where relative velocities are small;
on the other hand, ram--pressure stripping is only effective at very
high gas densities such as the cores of rich clusters. Therefore, by
comparing galaxy evolution in clusters of different environments it is
possible to separate the various physical processes that could
contribute to the observed transformation in the galaxy population of clusters.

\bigskip

{\it Acknowledgements:} I would like to thank my partners in Durham, Drs.
Bower, Davies, Kuntschner and Smail, as well as in Munich, Drs. Belloni, 
Bender, Greggio and Saglia, for the excellent collaboration, on which all
the presented results are based.

\vspace{0.7cm}
\noindent




\vfill

\end{document}